\documentclass[twocolumn,12pt,reprint]{aastex62}

\usepackage{epstopdf}
\usepackage{color}
\usepackage{graphicx}
\usepackage{booktabs}
\usepackage{nameref}
\usepackage{amsmath}

\newcommand{\goes}{GOES}
\newcommand{\chase}{CHASE}
\newcommand{\saas}{Sun-as-a-star}

\newcommand{\ha}{H$\alpha$}
\newcommand{\fei}{Fe {\small I}}
\newcommand{\ov}{O {\small V}}

\title{01_filament_sun_as_a_star}

\begin{document}
\title{\saas\ Analysis of the Solar Eruption Source Region Using \ha\ Spectroscopic Observations of CHASE}

\correspondingauthor{Yijun Hou}
\email{yijunhou@nao.cas.cn}

\author[0000-0002-3657-3172]{Xiaofeng Liu}
\affil{Key Laboratory of Dark Matter and Space Astronomy, Purple Mountain Observatory, Chinese Academy of Sciences, Nanjing 210023, China}
\affil{School of Astronomy and Space Science, University of Science and Technology of China, Hefei 230026, China}
\affil{State Key Laboratory of Solar Activity and Space Weather, National Astronomical Observatories, Chinese Academy of Science, Beijing 100101, China}

\author[0000-0002-9534-1638]{Yijun Hou}
\affil{State Key Laboratory of Solar Activity and Space Weather, National Astronomical Observatories, Chinese Academy of Science, Beijing 100101, China}
\affil{School of Astronomy and Space Science, University of Chinese Academy of Sciences, Beijing 100049, China}

\author[0000-0002-8258-4892]{Ying Li}
\affil{Key Laboratory of Dark Matter and Space Astronomy, Purple Mountain Observatory, Chinese Academy of Sciences, Nanjing 210023, China}
\affil{School of Astronomy and Space Science, University of Science and Technology of China, Hefei 230026, China}

\author[0000-0002-1190-0173]{Ye Qiu}
\affil{
Institute of Science and Technology for Deep Space Exploration, Suzhou Campus, Nanjing University, Suzhou 215163, China}

\author[0000-0001-6655-1743]{Ting Li}
\affil{State Key Laboratory of Solar Activity and Space Weather, National Astronomical Observatories, Chinese Academy of Science, Beijing 100101, China}
\affil{School of Astronomy and Space Science, University of Chinese Academy of Sciences, Beijing 100049, China}

\author[0009-0007-4469-0663]{Yingjie Cai}
\affil{State Key Laboratory of Solar Activity and Space Weather, National Astronomical Observatories, Chinese Academy of Science, Beijing 100101, China}
\affil{School of Astronomy and Space Science, University of Chinese Academy of Sciences, Beijing 100049, China}

\author[0000-0002-7544-6926]{Shihao Rao}
\affil{School of Astronomy and Space Science, Nanjing University, Nanjing 210023, China}
\affil{Key Laboratory of Modern Astronomy and Astrophysics (Nanjing University), Ministry of Education, Nanjing 210023, China}

\author{Junyi Zhang}
\affil{State Key Laboratory of Solar Activity and Space Weather, National Astronomical Observatories, Chinese Academy of Science, Beijing 100101, China}
\affil{School of Astronomy and Space Science, University of Chinese Academy of Sciences, Beijing 100049, China}

\author[0000-0001-7693-4908]{Chuan Li}
\affil{School of Astronomy and Space Science, Nanjing University, Nanjing 210023, China}
\affil{Key Laboratory of Modern Astronomy and Astrophysics (Nanjing University), Ministry of Education, Nanjing 210023, China}

\begin{abstract}
\saas\ analyses serve as a bridge for comparative studies on solar and stellar activities. To investigate the typical \saas\ \ha\ temporal spectral characteristics in solar eruption source regions, we analyzed five different types of solar eruptions, using spectroscopic data from the Chinese \ha\ Solar Explorer (CHASE). Because the spatially-integral \ha\ spectrum of source region is mainly contributed by emission from heated plasma in flare ribbons and absorption from cold plasma in evolving filaments, we separately analyze the sub-regions of the source region dominated by different dynamical processes. It is revealed that filament eruptions show emission near \ha\ line center, accompanied by blueshifted/redshifted absorption, while flare ribbons show \ha\ line center emission with red asymmetry and line broadening. Moreover, a special spectral signature likely associated with coronal mass ejections (CMEs) is identified: prominent blueshifted absorption without a clear deceleration phase, along with redshifted absorption, which can be used as a probe when searching stellar CMEs. Furthermore, in the X9.0 flare (SOL2024-10-03T12:18) accompanied by a violent CME, the expected blueshifted signal is not visible in the spatially-integral \ha\ spectra. This suggests that filament-lifting signals associated with CMEs in the source region can be obscured by the simultaneous dominant flare-ribbon emission within the integration region, which may explain the relatively small number of confirmed stellar CMEs observed in \ha. We also find that comparison between the \ha\ and UV spectral observations can effectively reveal the velocity evolution of erupting filaments and potential existence of associated CMEs.

\end{abstract}


\section{Introduction}\label{s1}

Solar flares are among the most powerful events on the Sun (e.g., \citealt{2011SSRv..159...19F}), appearing as sudden brightenings and releasing vast amounts of energy across a wide range of wavelengths. The released energy is transported to the lower atmosphere, heating the local plasma and causing it to move both upwards and downwards, known as chromospheric evaporation (\citealt{1982ApJ...263..409A}, \citealt{2006ApJ...638L.117M}, \citealt{2015ApJ...811....7L}) and chromospheric condensation (\citealt{1987SoPh..113..307F}, \citealt{1993ApJ...416..886G}, \citealt{2020ApJ...896..154Y}), respectively. Solar flares are usually accompanied by filament/prominence eruptions and coronal mass ejections (CMEs; e.g., \citealt{2011LRSP....8....1C}). There are three types of filament eruptions (\citealt{2007SoPh..245..287G}): full eruption, partial eruption (\citealt{2000ApJ...537..503G}, \citealt{2023ApJ...959...69H}), and failed eruption (\citealt{2003ApJ...595L.135J}, \citealt{2017ApJ...838...15L}). \citet{2003ApJ...586..562G} found that approximately $72\%$ of filament/prominence eruptions are associated with CMEs, which can interact with planetary magnetospheres and impact daily life (e.g., \citealt{2022LRSP...19....2C}).

Similar to the Sun, a large number of stellar flares have been observed on other stars, and many studies have compared them with solar flares \citep[e.g.,][]{2022ApJ...928..180W, 2022ApJ...933...92C, 2024LRSP...21....1K}. Statistical analyses have shown that the majority of stellar flares observed on solar-type stars are superflares, with energies exceeding $10^{33}$ erg (\citealt{2012Natur.485..478M}, \citealt{2013ApJS..209....5S}, \citealt{2024ApJ...961..130Z}). Detections of stellar filament/prominence eruptions and potential CMEs have also been reported based on Doppler shifts in the X-ray spectral observations and sudden dimmings in the extreme ultraviolet (EUV) and X-ray emissions (\citealt{2019NatAs...3..742A}, \citealt{2021NatAs...5..697V}). Understanding the properties of stellar CMEs is important for assessing their potential impact on both stellar evolution and exoplanet habitability \citep{2024ApJ...961..189N}. For example, stellar CMEs can contribute to mass and angular momentum loss, thereby influencing the star’s long-term evolution \citep[e.g.,][]{2015ApJ...809...79O, 2017MNRAS.472..876O, 2021ApJ...915...37W}. In addition, frequent or energetic CMEs can significantly affect exoplanetary environments by altering atmospheric composition and increasing radiation exposure \citep[e.g.,][]{2007AsBio...7..185L, 2016NatGe...9..452A, 2021NatAs...5..298C}.

Using \ha\ spectral observations, \cite{2022NatAs...6..241N} detected a filament eruption on the solar-type star EK Draconis, with an estimated mass of $1.1^{+4.2}_{-0.9}\times 10^{18}$ g, accompanied by a CME exhibiting a line-of-sight velocity of $\sim$$-$510 km s$^{-1}$. On the same solar-type star, \cite{2024ApJ...961...23N} further reported the discovery of two prominence eruptions, observed as blueshifted \ha\ emissions at speeds of 690 and 430 km s$^{-1}$ and masses of $1.1 \times 10^{19}$ and $3.2 \times 10^{17}$ g, respectively. In the case of M-dwarfs, many studies in recent years have reported possible filament/prominence eruptions by identifying blue asymmetries in chromospheric line profiles \citep[e.g.,][]{2016A&A...590A..11V, 2024ApJ...961..189N, 2025ApJ...978L..32L, 2025ApJ...979...93K, 2025ApJ...985..136K}. \cite{2024ApJ...961..189N} reported 7 out of 41 \ha\ flare events showing blue wing asymmetries, with velocities ranging from $-$73 to $-$122 km s$^{-1}$, and the mass of upward-moving plasma ranging from $\sim$10$^{15}$ to 10$^{19}$ g. \cite{2025ApJ...978L..32L} reported a stellar prominence eruption, with a mass range from $1.6 \times 10^{19}$ to $7.2 \times 10^{19}$ g, accompanied by a CME exhibiting an extreme maximum blueshifted velocity of $-$605 $\pm$ 15 km s$^{-1}$. It is worth noting here that the filament eruptions can also be observed in emission on M-dwarfs (\citealt{2022MNRAS.513.6058L}), which is different from the absorption signals observed in solar and G-type stars, and may bring some specific difficulties when interpret the corresponding physical processes. To explore the mechanism of CMEs in active stars, \cite{2018ApJ...862...93A} presented a 3D Magnetohydrodynamics numerical simulation and found that greater energy is required for plasma to escape under stronger magnetic fields.

Despite extensive studies in recent years, due to the limited stellar observations, the detailed physical mechanisms of stellar eruptive events are not clearly clarified, leaving many urgent questions to be addressed \citep[e.g.,][]{2024ApJ...961..189N, 2024ApJ...964...75O, 2024A&A...682A..46P}. For example, existing observations indicate that stellar flares are common (\citealt{2023A&A...669A..15Y}) while reliable stellar CMEs are relatively rare. Is this disparity due to strong stellar magnetic fields preventing CME formation, or are there other underlying reasons related to observational effect? Additionally, do special plasma dynamics processes, such as failed and partial filament eruptions, also occur in the source regions of stellar eruptions? Directly answering these questions requires time-series stellar observations with spatial resolution, which remains challenging to achieve so far.

Unlike the other distant stars, the Sun, as the closest star to Earth, can be observed in detail with high spatial and temporal resolutions. In recent years, spatially-integral solar observed data have been widely used to compare with and interpret stellar observations, i.e., a method known as `\saas\ analysis' (\citealt{2016SoPh..291.1761H}). In recent studies, researchers have identified characteristic features of \saas\ \ha\ line profiles for flares, filament/prominence eruptions, and CMEs (\citealt{2022ApJ...933..209N}, \citealt{2022ApJ...939...98O}). By varying the spatial size of the integration region, \cite{2024ApJ...966...45M} used \ha\ spectroscopic data from the Chinese \ha\ Solar Explore (\chase; \citealt{2022SCPMA..6589602L}) and found that the redshift velocity derived from the spatially-integral \ha\ profile remains unchanged as the size of the integration region increases. There are also some works using \saas\ spectroscopic observations at EUV wavelengths \citep[e.g.,][]{2022ApJ...931...76X, 2023ApJ...953...68L, 2024ApJ...964...75O}. Particularly, \cite{2024ApJ...964...75O} detected \saas\ line profile blueshifts in both \ha\ and \ov\ 629.7 \AA\ wavebands, associated with the filament eruption and CME processes during an M8.7 flare event. Additionally, simulation studies have been conducted to reproduce observed spectral lines and provide insights into local environmental parameters (\citealt{2024ApJ...963...50I}, \citealt{2025A&A...694A.315Y}).

Although the \saas\ analysis method can identify key features of major solar eruptions, it has some inherent flaws. It is well known that multiple physical processes occur simultaneously in the source region during a solar eruption, collectively shaping the overall \saas\ spectral characteristics \citep[e.g.,][]{2019ApJ...875...93C, 2022ApJ...933..209N, 2024ApJ...974L..13O}. Therefore, it is difficult to accurately extract the typical spectral signatures of different physical processes from the spatially-integral spectra of the entire solar disk. To better distinguish the spectral features of different physical processes in the solar eruption source region, a more detailed analysis using sub-region \saas\ studies is necessary.

In this paper, we analyze \ha\ spectroscopic observations of different types of solar eruptions and perform virtual \saas\ analysis on sub-regions dominated by different physical processes within the source region. Our goal is to identify the spectral characteristics associated with various dynamical processes, such as chromospheric condensation in flare ribbons, the upward and downward motion of cold plasma within filaments, and their interactions. Additionally, we conduct a comparative analysis of flares accompanied by full, partial, and failed filament eruptions, trying to identify more reliable spectral indicators of CMEs through \saas\ spectra. In Section \ref{s2}, we describe the instruments and observational data. Section \ref{s3} outlines the analytical method. In Section \ref{s4}, we present and discuss the results. Finally, Section \ref{s5} provides a summary of our findings.

\section{Observations}\label{s2}
The \ha\ Imaging Spectrograph (HIS) onboard \chase\  provides full-disk spectroscopic observations of the Sun in \ha\ (6559.7--6565.9 \AA) and \fei\ (6567.8--6570.6 \AA) wavebands. Considering the relative motion between the satellite and the Sun, and the additional influence of solar rotation, we use the wavelength range of \ha\ line center $-$3.3--$+$2.2 \AA\ for our analysis, which corresponds to a Doppler velocity of approximately $-$150--$+$100 km s$^{-1}$. However, in some solar eruptions investigated here, the observed blueshift signals could reach a maximum velocity exceeding $-$150 km s$^{-1}$, implying that the corresponding velocity estimation can be limited by the CHASE wavelength coverage. In Section~\ref{s4}, we discuss the potential impact of this constraint on the results of each affected event in detail. The \chase 's data used in our study have been corrected, including slit-image-curvature, dark-field, and flat-field corrections (\citealt{2022SCPMA..6589603Q}). They have been binned to $\sim$1\arcsec\ pixel$^{-1}$ and $\sim$0.048 \AA\ pixel$^{-1}$ in spatial and spectral resolutions, respectively, and the temporal resolution is $\sim$71 s. The Level 1 \chase\ data are available through the website of Solar Science Data Center of Nanjing University (SSDC)  \footnote[1]{\href{https://ssdc.nju.edu.cn}{https://ssdc.nju.edu.cn}}. We redo the wavelength calibration using Gaussian fitting of the line profiles in quiet-Sun area near the target region, which can remove the effect of solar rotation. The resultant calibration accuracy is approximately 0.02 \AA , which corresponds to about 1 km s$^{-1}$ in Doppler velocity.

We also use the soft X-ray (SXR) flux at 1–8 Å observed by the X-Ray Sensor (XRS) on the Geostationary Operational Environmental Satellite (GOES) and the data of the Large Angle and Spectrometric Coronagraph (LASCO; \citealt{1995SoPh..162..357B}) on board the Solar and Heliospheric Observatory (SOHO) to determine the occurrence of flares and CMEs, respectively. Additionally, we use the imaging data of 131 and 171 \AA\ taken by the Atmospheric Imaging Assembly (AIA; \citealt{2012SoPh..275...17L}), and the spectral observations taken by the Extreme ultraviolet Variability Experiment (EVE; \citealt{2012SoPh..275..115W}), both onboard the Solar Dynamics Observatory (SDO; \citealt{2012SoPh..275....3P}), for studying evolution of  the plasma structure deriving from eruption source region in hotter EUV wavelengths.

We have selected 5 solar eruptive events, with relatively complete \chase\ observations, to identify the typical \saas\ spectral features of different physical processes occurring in the solar eruption source regions. The events include flare without obvious filament eruption (Event 1), quiescent filament eruption without remarkable flare ribbons in quiet region (Event 2), flares with successful filament eruption (Events 3 \& 4), and failed filament eruption (Event 5). Part of the information of these events are shown in Table \ref{tab:1}, and the details are given in Section \ref{s4}.

\begin{table*}[htbp]

    \centering
    \caption{Information of the events analyzed in this work}
    \begin{tabular}{c c c c c c c}
        \toprule
        \fontsize{10}{12}
        Event & Date (UT) & GOES peak time (UT) & GOES Class & Location & Event Features & CME? \\
        \midrule
        1 & 2024 Jun 16 & 15:46 & C6.7 & S26E05 & flare without obvious filament eruption & No \\
        2 & 2024 Apr 11 & 05:30 & B8.8 & N22E09 & quiescent filament eruption & Yes \\
         & & & & & without remarkable flare ribbons & \\
        3 & 2024 Jul 29 & 12:55 & M8.7 & S11W50 & flare with successful filament eruption & Yes \\
        4 & 2024 Oct 03 & 12:18 & X9.0 & S15W03 & flare with successful filament eruption & Yes \\
        5 & 2024 Aug 20 & 16:39 & C4.9 & N23E13 & flare with failed filament eruption & No \\
        \bottomrule
    \end{tabular}

    \label{tab:1}
\end{table*}

\section{Method}\label{s3}
\subsection{\saas\ Analysis}\label{s31}
The change amplitude of \ha\ line profile caused by solar flares occurring in local region is much smaller than the background full-disk solar radiation ($\sim 10^{-4}-10^{-5}$), which is almost negligible in the \saas\ analysis. Although there are a few works trying the `true \saas\ method' (e.g., \citealt{2024ApJ...966...45M}), the signal-to-noise ratio (SNR) is still quite low. This work aims to deconstruct different sub-regions' contribution to the \saas\ line profile, and figure out the differences of line profiles between events with CME or not, which needs relatively high SNR. As a result, we use `virtual \saas\ method' (\citealt{2022ApJ...939...98O}), that is, converting the spatially-integral \ha\ spectra of target regions (TRs) to \saas\ spectra, with an assumption that the variations of \saas\ line profiles are only caused by TRs. The method's details are as follows:

First, we calculate the spatially-integral \ha\ spectra $f$ inside region $A$:

\begin{equation}
    f(t,\lambda,A)=\int_{A}I(t,\lambda,x,y)dxdy,
\end{equation}
where $I(t, \lambda, x, y)$ is the intensity function of the observed time $t$, the wavelength $\lambda$, and the spatial position $(x,y)$. Then we do normalization to suppress the fluctuations from instrument, using continuum spectra level at each time:

\begin{equation}
    F(t,\lambda) = \frac{f(t,\lambda,A=TR)}{f(t,\lambda_{cont},A=TR)} \times f(t_{0},\lambda_{cont},A=TR),
\end{equation}
where $t_{0}$ represents a pre-event time and $\lambda_{cont}$ is a continuum wavelength in the spectral window of \fei . Finally, we obtain the pre-flare subtracted \saas\ \ha\ line profiles, normalized by the full-disk continuum level:

\begin{equation}
    \Delta S(t,\lambda) = \frac{F(t,\lambda) - F(t_0,\lambda)}{f(t_{0},\lambda_{cont},A=full \ disk)}.
\end{equation}

To reveal the total change of the \ha\ spectrum, we also get the differenced equivalent width:

\begin{equation}
    \Delta EW = \int_{H\alpha - \Delta \lambda}^{H\alpha + \Delta \lambda}\Delta S(t,\lambda)~d\lambda,
\end{equation}
where $H\alpha$ represents the line center of \ha\ line. To better utilize the CHASE instrument’s wavelength coverage and to comprehensively capture the influence of blueshifted absorption on the overall spectra, we set the integration range $\Delta\lambda$ from $-$3 \AA\ to $+$2 \AA\ relative to the \ha\ line center.

\subsection{Doppler Velocity Calculation}\label{s32}
To quantitatively analyze dynamical processes in different sub-regions of the eruption source region, we apply different methods to calculate Doppler velocity distribution for each region through \chase\ \ha\ observations. In quiet Sun and flare ribbon regions, we use moment analysis (\citealt{2020ApJ...896..154Y}), while in filament regions we use single- and two-cloud model (\citealt{2024ApJ...961L..30Q}). Moment analysis calculates the wavelength centroid of a given line profile ($\lambda_{c}$) and compares it with the reference wavelength center ($\lambda_{r}$) to determine the velocity parameter. The specific calculation processes are as follows:

\begin{equation}
    \lambda_{c}=\frac{\int \lambda(I(\lambda)-I_{0})d\lambda}{\int (I(\lambda)-I_{0})d\lambda}.
\end{equation}
In quiet Sun region, $I(\lambda)$ is the observed intensity of original line profiles, while in flare ribbon region, $I(\lambda)$ is the intensity of pre-flare subtracted spectra. $I_{0}$ is the continuum background intensity of $I(\lambda)$. Then we calculate the velocity by:

\begin{equation}
    v=c(\lambda_{c}-\lambda_{r})/\lambda_{r},
\end{equation}
where $c$ is the light speed. The cloud model assumes a cold, cloud-like plasma suspended above a hotter local environment. In single-cloud model, the intensity of the filament can be calculated using the radiative transfer equation:

\begin{equation}
    I=C_{ld}I_{0}e^{-\tau}+S(1-e^{-\tau}),
\end{equation}
where $C_{ld}$ is a parameter representing the influence of limb darkening, $I_{0}$ denotes the background intensity, and $S$ is the source function of the filament. $\tau$ represents the optical depth, which can be calculated as follows:

\begin{equation}
    \tau(\lambda)=\tau_{0}exp[-(\lambda-\lambda_{0}-v\lambda_{0}/c)^{2}/W^{2}],
\end{equation}
where $v$ represents the Doppler velocity and $W$ denotes the line width. We can obtain the value of $v$, $C_{ld}$, $S$, $\tau_{0}$ and $W$ by fitting the observed \ha\ line profiles. When the filament plasma becomes more complex and the single-cloud model cannot accurately fit the line profiles, we apply the two-cloud model:

\begin{equation}
    I=C_{ld}I_{0}e^{-(\tau_{1}+\tau_{2})}+S_{1}(1-e^{-\tau_{1}})e^{-\tau_{2}}+S_{2}(1-e^{-\tau_{2}}),
\end{equation}
which can get the parameters for both upwards and downwards moving components.

\subsection{Some Other Spectral Parameter Calculations}\label{s33}

To measure the line broadening width and redshifted velocity of the pre-flare subtracted \ha\ spectra ($\Delta S(t,\lambda)$, as described in Section \ref{s31}) within the solar eruption source regions, contributed by hot plasma in flare ribbons, we adopt a two-component fitting approach, which is usually used both in stellar (\citealt{2022ApJ...928..180W}) and \saas\ (\citealt{2022ApJ...933..209N}; \citealt{2024ApJ...966...45M}) observations. The fitting model consists of a central Voigt function and a Gaussian component on red side. The line broadening width is defined as the width of one tenth maximum of the central Voigt function, while the redshifted velocity is obtained from the centroid of the red-wing Gaussian component.

For each event with available EVE data, the pre-event spectrum is calculated as the time-averaged spectrum over the 18-minute interval preceding the start of each \chase\ observation. The reference wavelength center is then determined by performing a single Gaussian fit to the pre-event spectrum of each event. Subsequently, the per-event-subtracted spectra are obtained by subtracting the pre-event spectrum from the original spectra and normalizing the result by the peak irradiance of the pre-event spectrum. We calculate the noise-induced fluctuation using all pre-event spectra data. At each wavelength point, we compute the standard deviation of irradiance variations along time. The fluctuation, $\sigma_{max}$, is defined as the maximum of these standard deviations across all wavelengths. These are similar to the methods used in \cite{2024ApJ...964...75O}.

To estimate the blueshifted velocity of the pre-event subtracted EVE line profiles, we apply a double Gaussian fitting method, similar to the approach in \cite{2022ApJ...931...76X}. The model includes two Gaussian components representing the central and blue-wing features. In Event 4, we observed a redshifted velocity exceeding $+$50 km s$^{-1}$, therefore, we allowed the central wavelength to vary over a wider range, from $-$100 to $+$100 km s$^{-1}$. The blueshifted velocity is derived from the centroid of the blue-wing Gaussian component. In this study, we focus on the \ov\ 629.7 \AA\ line from EVE, as it provides clearer signatures and allows for the detection of higher velocity components (\citealt{2024ApJ...964...75O}).

\begin{figure*}[htbp]
    \centering
    \includegraphics[width=0.9\textwidth]{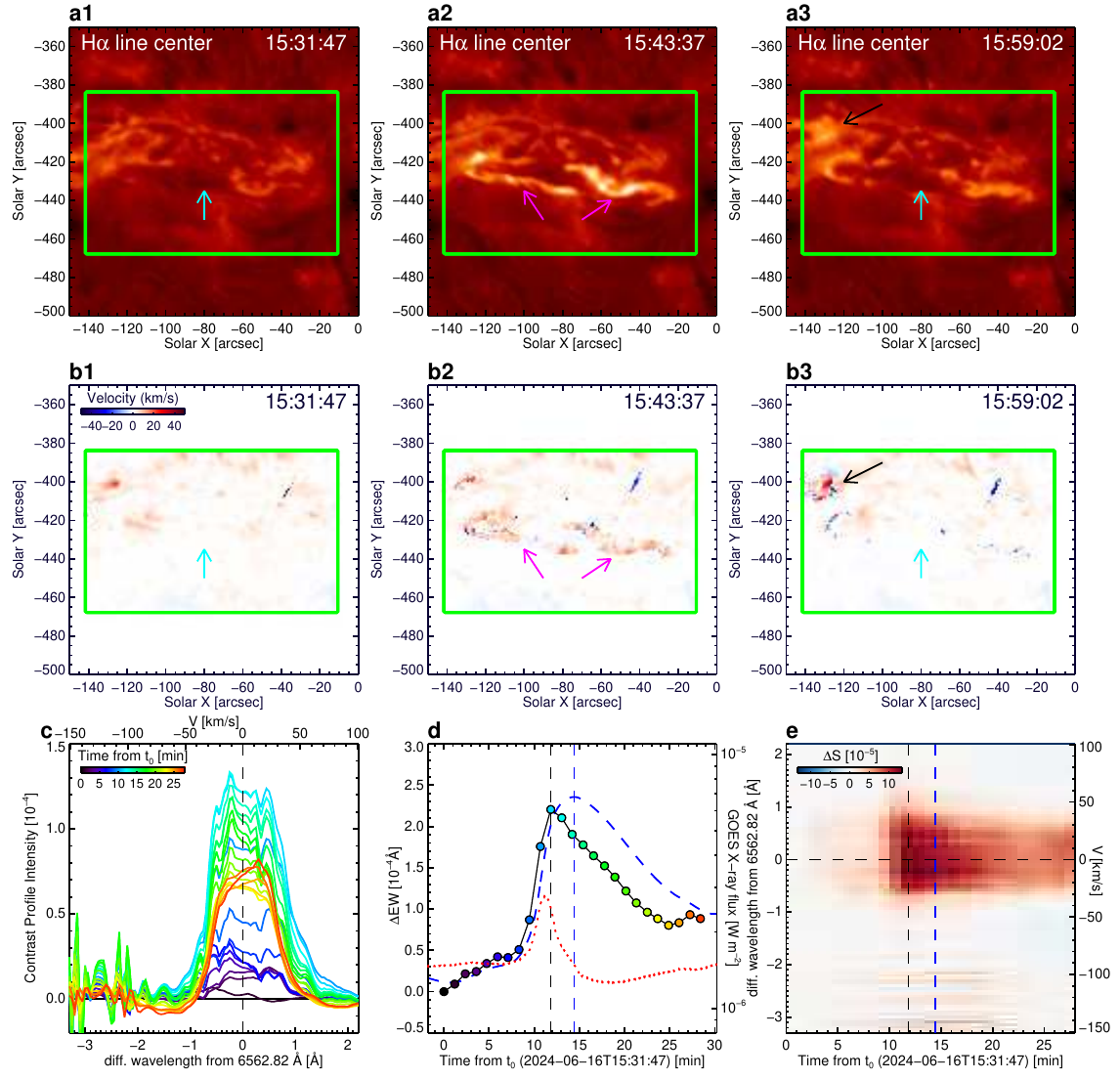}
    \caption{\chase~imaging and spectral observations of the C6.7 flare, occurred on 2024 June 16. (a1)--(a3): Sequence of CHASE \ha\ line center images showing the source region (green boxes). The cyan, magenta, and black arrows respectively mark the filament, flare ribbons, and brightening related to another flare. (b1)--(b3): Doppler velocity fields of the source region, derived from \chase~\ha\ spectra.  (c): Evolution of contrast Sun-as-a-star \ha\ line profiles. (d): Light curves of \ha\ differential equivalent widths (colored circles), GOES soft X-ray 1--8~\AA~flux (blue dashed curve) and its temporal derivative (red dotted curve). Vertical dashed lines denote the \ha\ (black) and GOES (blue) peak times. (e): The time series of Sun-as-a-star \ha\ dynamic spectrum in the source region. An animation (Figure1.mp4) covering 15:31 UT to 16:00 UT is available online, which displays the evolution of Event 1 through \ha\ $-$ 1 \AA\ , \ha\ center, and \ha\ $+$ 1 \AA\ intensity images. The animation’s duration is 3 seconds.}
    \label{fig_1}
\end{figure*}

\begin{figure*}[htbp]
	\centering
	\includegraphics[width=0.9\textwidth]{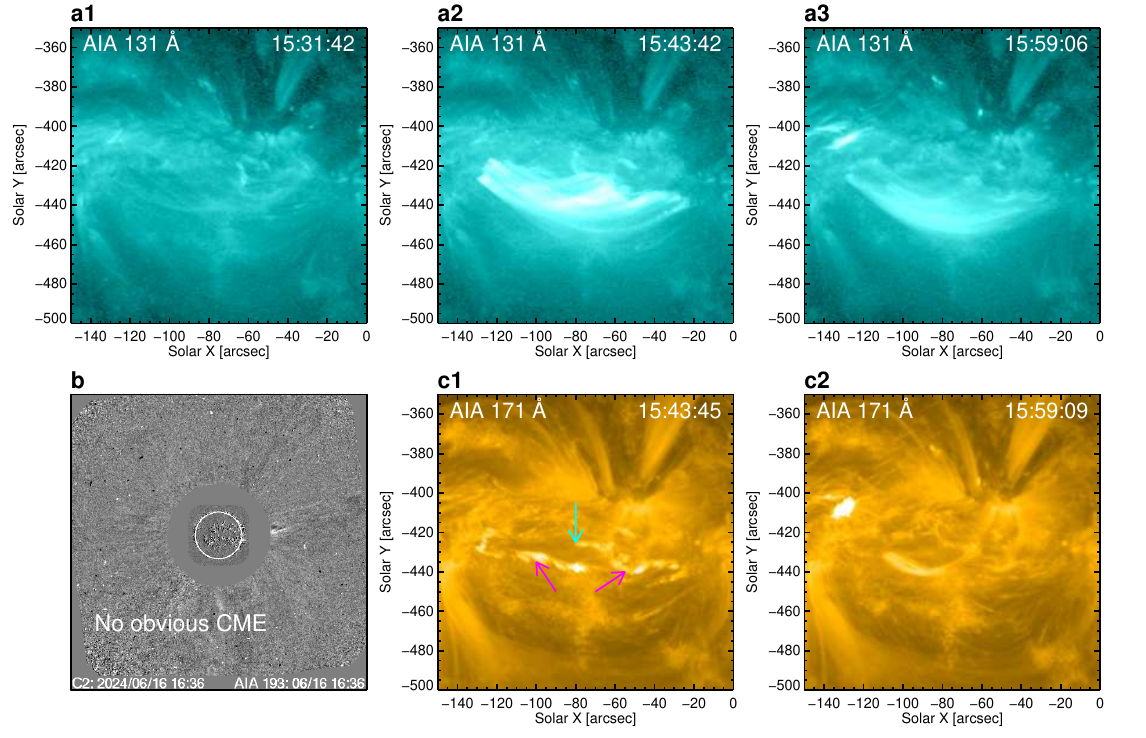}
	\caption{Multi-wavelength observations of the C6.7 flare that occurred on 2024 June 16. (a1)--(a3): AIA 131~\AA\ images taken at times corresponding approximately to those shown in panels (a1)--(a3) of Figure~\ref{fig_1}. (b): LASCO C2 running difference image at 16:36 UT, retrieved from \href{https://cdaw.gsfc.nasa.gov/}{https://cdaw.gsfc.nasa.gov/}. (c1)--(c2): AIA 171~\AA\ images taken at times corresponding approximately to those shown in panels (a2)--(a3) of Figure~\ref{fig_1}. The cyan and magenta arrows in panel (c1) indicate the filament and flare ribbons, respectively. }
	\label{fig_a1}
\end{figure*}

\section{Results and Discussion}\label{s4}
Previous studies have shown that the spatially-integral spectral characteristics of solar eruption source regions were mainly influenced by two contributing factors: emission from heated plasma in flare ribbons and absorption from cold plasma in evolving filaments. In this study, we investigate five solar eruption events from simple to complex. The \saas\ temporal spectral characteristics of Events 1 and 2 are dominated by a single contributing factor, whereas those of Events 3, 4, and 5 result from the combined influence of both factors. To better disentangle these effects, we perform separate analyses of sub-regions within the source region. These sub-regions are identified and isolated based on the imaging observations and Doppler velocity map, with each region dominated by a distinct dynamical process. By conducting spatially-integral spectra analysis on these sub-regions, we then elucidate typical temporal spectral characteristics of these different dynamical processes.

\subsection{Solar Eruptions with Spectral Characteristics Dominated by a Single Contributing Factor}

\subsubsection{Flare without Obvious Filament Eruption}
Figure \ref{fig_1} presents the imaging and spectral observations of Event 1, a C6.7 flare occurred on 2024 June 16 (see the online animation associated with Figure \ref{fig_1}). This flare is classified as a type-I confined flare (according to the result in \citealt{2019ApJ...881..151L}), which is accompanied by no obvious CME (see Figure \ref{fig_a1}(b)). Through \ha\ imaging observations (Figures \ref{fig_1}(a1)--(a3)), the filament located above the flare ribbons remains stable throughout this event. Consistently, no eruptive structures are observed in AIA 131 and 171 \AA\ images (see Figures \ref{fig_a1}(a1)--(a3), (c1), and (c2)). As a result, the spatially-integral spectral characteristics of the source region is dominated by the emission from flare ribbons (marked by magenta arrows in Figures \ref{fig_1}(a2) and (b2)). According to GOES observations, the flare began at 15:34 UT, peaked at 15:46 UT, and ended at 15:54 UT. The \saas\ spectra (shown in Figures \ref{fig_1}(c) and (e)) initially display roughly symmetrical profiles with increasing line center emission and line broadening. Near the \ha\ peak time, the spectral profiles exhibit line broadening over about $\pm$1.4 \AA, and red asymmetry with redshift velocity of $\sim$$+$30 km s$^{-1}$, consistent with previous findings and typically interpreted as a result of chromospheric condensation (\citealt{2022ApJ...933..209N}). During the decay phase, the emission intensity gradually diminishes, and the line profiles retain a symmetrical and single-peaked structure. The $\Delta EW$ shows a significant enhancement, peaking during the rise phase of the flare, before the peak of the GOES SXR flux and near its temporal derivative (see Figure \ref{fig_1}(d)). It should be noted that, the temporal resolution of \chase\ observation is limited, which may cause inaccuracies when determining the precise peak time of $\Delta EW$. The limitation on wavelength coverage of CHASE spectral observation has minimal impact on the analysis of this event, because all the line broadening or asymmetry signals are within the wavelength range. The fringe-like noise signals in the blue wing are caused by the Si I and Ti II lines, which lie outside the wavelength range of our target and exhibit relatively weak intensities, and they do not affect our main results. Additionally, a secondary peak appears near 16:00 UT, attributed to the brightening in the northeastern part of the source region (marked by the black arrow in Figures \ref{fig_1} (a3) and (b3)).

\begin{figure*}[htbp]
    \centering
    \includegraphics[width=0.9\textwidth]{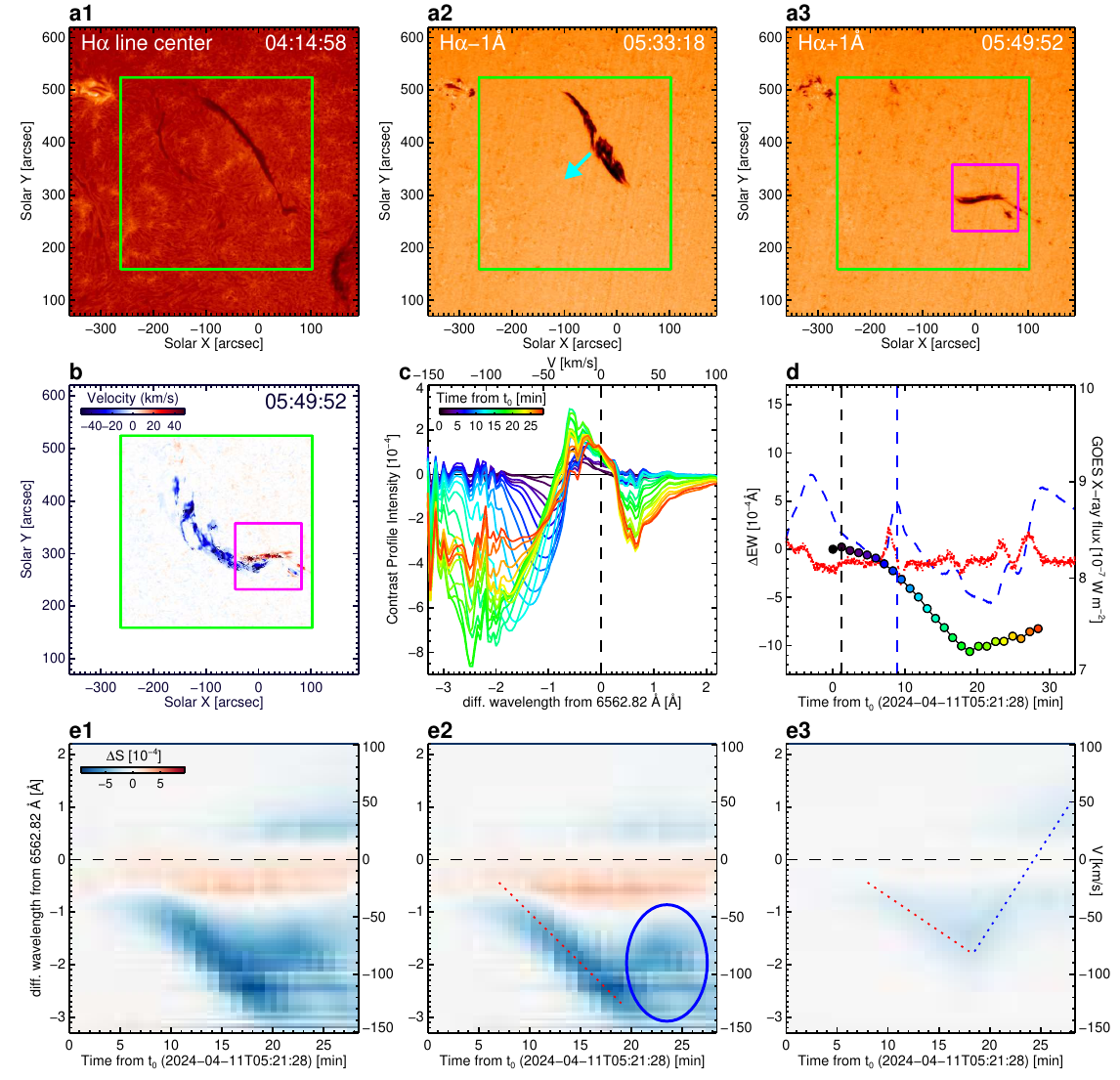}
    \caption{\chase~imaging and spectral observations of the quiescent filament eruption event, occurred on 2024 April 11. (a1)--(a3): Sequence of CHASE \ha\ line center, blue wing, and red wing images showing the source region (green boxes). The cyan arrow indicates the erupting direction. The magenta box marks the region with redshifted velocity. (b): Doppler velocity fields of the source region at the same time shown in (a3), derived from \chase~\ha\ spectra. (c)--(d): The same as Figures \ref{fig_1} (c)--(d) , but for this event. (e1)--(e3): The time series of Sun-as-a-star \ha\ dynamic spectrum inside the green box, inside the green box but outside the magenta box, and inside the magenta box, respectively. An animation (Figure3.mp4) covering 04:10 UT to 04:14 UT and 05:21 UT to 05:49 UT is available online, which displays the evolution of Event 2 through \ha\ $-$ 1 \AA\ , \ha\ center, and \ha\ $+$ 1 \AA\ intensity images. The animation’s duration is 3 seconds.}
    \label{fig_2}
\end{figure*}

\begin{figure*}[htbp]
	\centering
	\includegraphics[width=0.9\textwidth]{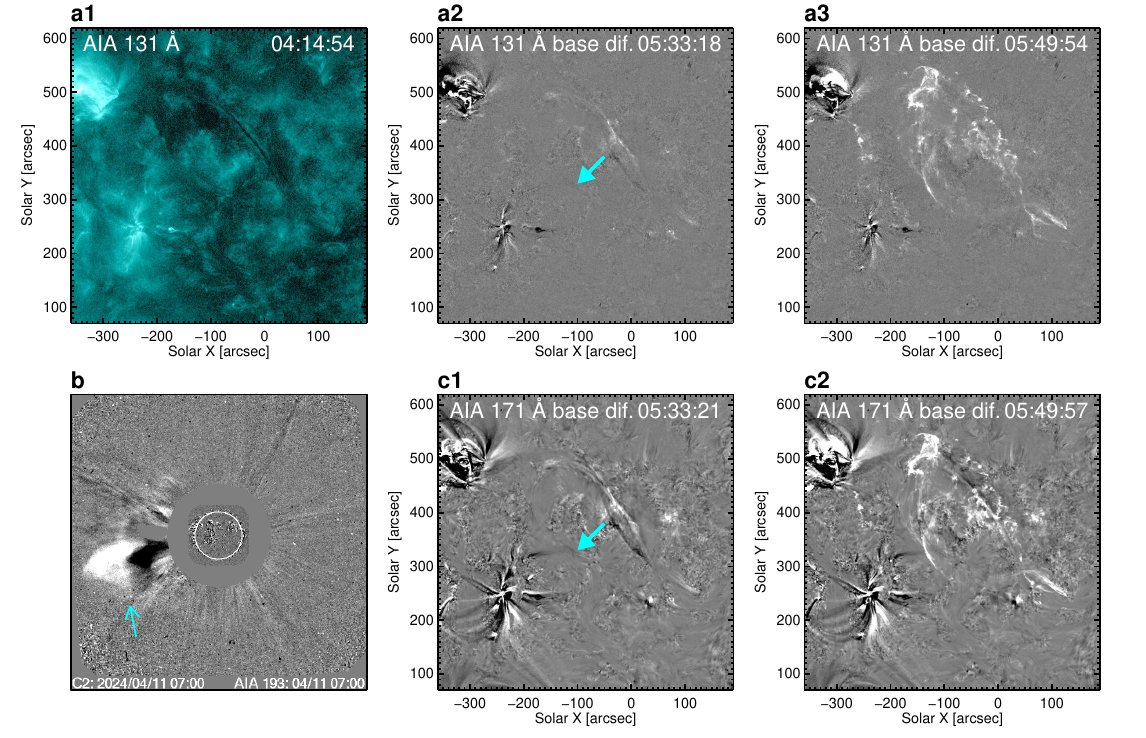}
	\caption{Multi-wavelength observations of the quiescent filament eruption that occurred on 2024 April 11. (a1)--(a3): AIA 131~\AA\ raw and running difference images taken at times corresponding approximately to those shown in panels (a1)--(a3) of Figure~\ref{fig_2}. (b): LASCO C2 running difference image at 07:00 UT, with a cyan arrow indicating the CME associated with this filament eruption event. (c1)--(c2): AIA 171~\AA\ running difference images taken at times corresponding approximately to those shown in panels (a2)--(a3) of Figure~\ref{fig_2}. The cyan arrows in panels (a2) and (c1) indicate the direction of filament eruption. }
	\label{fig_a2}
\end{figure*}

\subsubsection{Quiescent Filament Eruption without Remarkable Flare Ribbons}
Figure \ref{fig_2} presents the imaging and spectral observations of Event 2, a quiet-Sun filament eruption occurred on 2024 April 11 in quiet Sun region, accompanied by a B8.8 class flare (see the online animation associated with Figure \ref{fig_2}). In this event, there are no remarkable flare ribbons and the radiation from flare ribbon regions is relatively low, resulting in spatially-integral spectral characteristics dominated by the absorption from the filament's moving plasma. The GOES light curve exhibits an enhancement during the event, whereas $\Delta EW$ shows a decline (see Figure \ref{fig_2}(d)). This is a partial eruption event, with most of the filament structure erupts successfully and causes a CME (see Figure \ref{fig_a2}(b)), while part of the plasma falls back near the footpoints (as outlined by the magenta box in Figures \ref{fig_2}(a3) and (b)). In AIA 131 and 171 \AA\ imaging observations, the filament eruption structures are visible as bright features in base difference images (Figures \ref{fig_a2}(a2), (a3), (c1), and (c2)). To distinguish the different \saas\ \ha\ spectral features associated with full, failed, and partial eruptions, we separately analyze regions without redshift velocities, with redshift velocities, and the combination of both. We assume that these three sub-regions represent idealized cases of full, failed, and partial filament eruptions, respectively. Specifically, the first region only includes plasma with continuous upward motion (full eruption), the second region includes plasma that first rises and then falls back (failed eruption), and the third region contains both continuously rising and partially descending plasma (partial eruption). Figures \ref{fig_2}(e1)–(e3) show the typical time evolution of \saas\ spectra for the partial, full, and failed eruption regions, respectively.

By analyzing the \saas\ spectra, we find that both the full and failed eruption regions exhibit accelerated blueshifted followed by decelerated blueshifted absorption, as well as emission near the line center. This emission results from the combined effects of filament disappearance and weak flare brightening, which is also mentioned by \cite{2022ApJ...939...98O}. However, there are two key differences between the two regions: (1) Acceleration phase: The peak Doppler velocity of plasma in the full eruption region is higher ($\sim$$-$130–-140 km s$^{-1}$, see red dotted line in Figure \ref{fig_2}(e2)) compared to that in the failed eruption region ($\sim$$-$80--90 km s$^{-1}$, see red dotted line in Figure \ref{fig_2}(e3)). (2) Deceleration Phase: In the failed eruption region, the blueshifted absorption component decelerates and turns into redshifted absorption (see the blue ellipse in Figure \ref{fig_2}(e3)), indicating that the filament plasma changes moving direction from upward to downward. In contrast, in the full eruption region, the blueshifted absorption is relatively strong and does not turn into redshifted absorption within the period of CHASE observation (see the blue ellipse in Figure \ref{fig_2}(e2)). In the partial eruption region (Figure \ref{fig_2}(e1)), the \saas\ spectra exhibit characteristics of both full and failed eruption regions. It is obvious that the blueshifted absorption signals in all the three regions are within the wavelength range and the differences between the full and failed eruption regions remain clearly distinguishable. Therefore, the impact of the CHASE wavelength coverage limitation is minimal here.

In this event, the proportion of the full eruption region is larger than that of the failed eruption region, causing the \saas\ \ha\ spectral features of the combined region to be dominated by those of the full eruption. As a result, the absorption strength of the redshifted component is significantly weaker than that of the blueshifted component. These findings indicate that the presence of blue- or redshifted absorption in \saas\ \ha\ spectra cannot serve as conclusive evidence for the existence of CMEs. This highlights a key challenge in detecting stellar CMEs using spectral methods, as similar spectral signatures may arise from different underlying physical processes.

\begin{figure*}[htbp]
    \centering
    \includegraphics[width=0.9\textwidth]{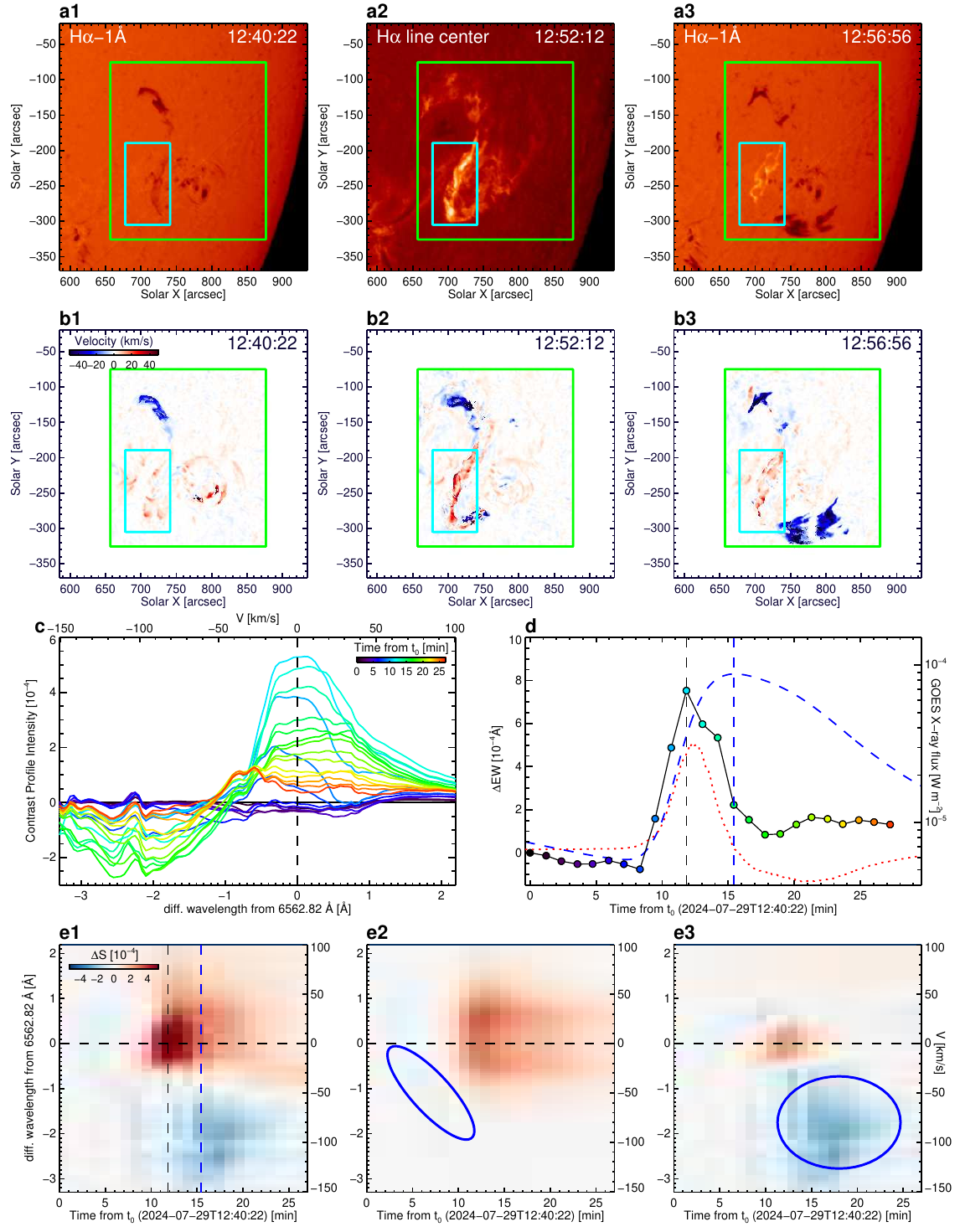}
    \caption{\chase~imaging and spectral observations of the M8.7 flare, occurred on 2024 July 29. (a1)--(a3): Sequence of CHASE \ha\ blue wing and line center images showing the source region (green boxes). The cyan box marks the brightening region of flare ribbon. (b1)--(d): The same as Figures \ref{fig_1} (b1)--(d) , but for this event. (e1)--(e3): The time series of Sun-as-a-star \ha\ dynamic spectrum inside the green box, inside the cyan box, and inside the green box but outside the cyan box, respectively. An animation (Figure5.mp4) covering 12:40 UT to 13:07 UT is available online, which displays the evolution of Event 3 through \ha\ $-$ 1 \AA\ , \ha\ center, and \ha\ $+$ 1 \AA\ intensity images. The animation’s duration is 3 seconds.}
    \label{fig_3}
\end{figure*}

\begin{figure*}[htbp]
	\centering
	\includegraphics[width=0.9\textwidth]{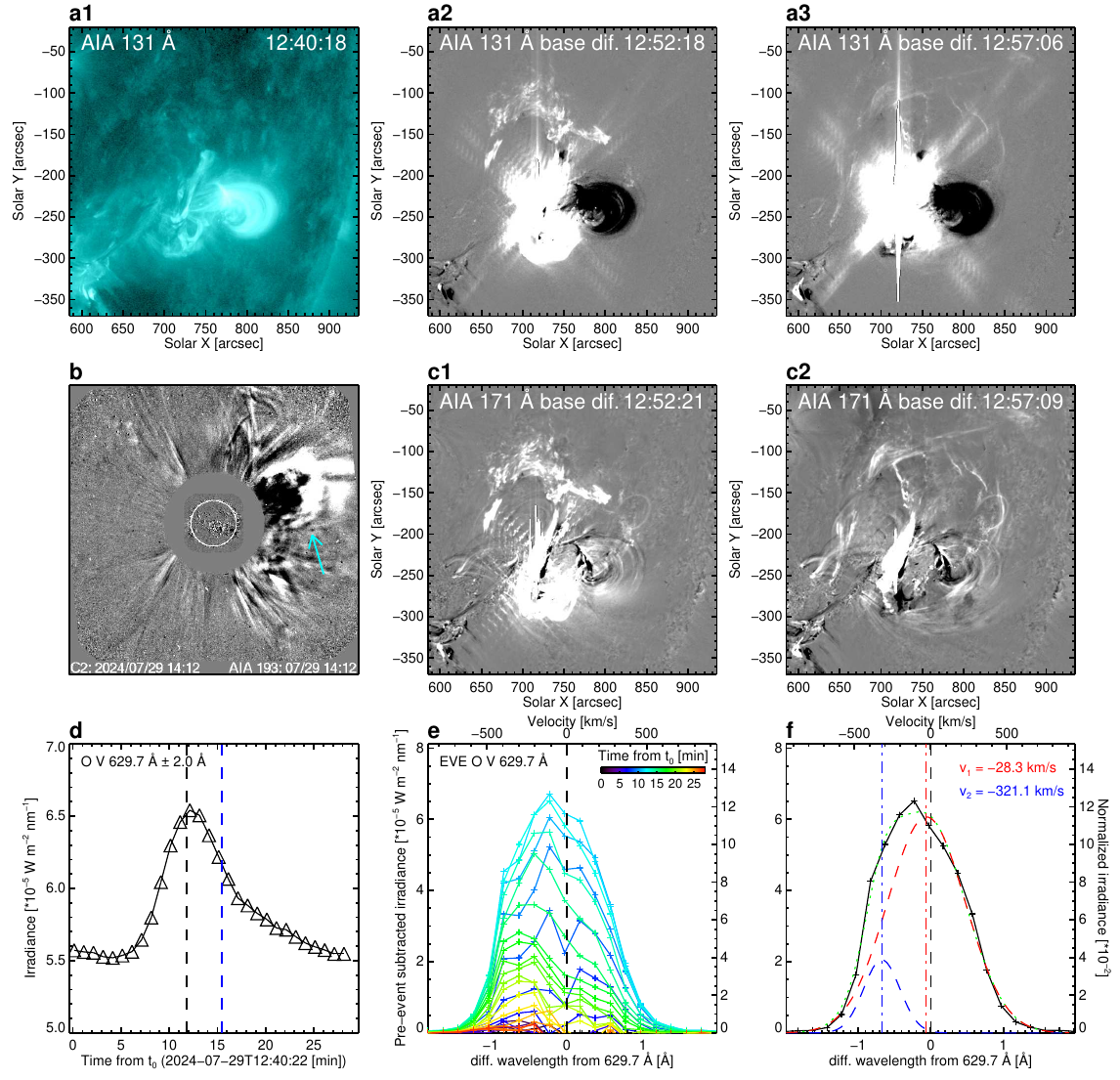}
	\caption{Multi-wavelength observations of the M8.7 flare, occurred on 2024 July 29. (a1)--(a3): AIA 131~\AA\ raw and running difference images taken at times corresponding approximately to those shown in panels (a1)--(a3) of Figure~\ref{fig_3}. (b): LASCO C2 running difference image at 14:12 UT, with a cyan arrow indicating the CME associated with this filament eruption event. (c1)--(c2): AIA 171~\AA\ running difference images taken at times corresponding approximately to those shown in panels (a2)--(a3) of Figure~\ref{fig_3}. (d): Irradiance variations of \ov\ 629.7 \AA. The vertical dashed lines mark the \ha\ (black) and GOES (blue) peak times. (e): Evolution of pre-event-subtracted O~V spectra during this event. (f): Double Gaussian fitting result of the \ov\ line profile near its peak time, showing distinct central and blueshifted emission components.  }
	\label{fig_a3}
\end{figure*}

\subsection{Solar Eruptions with Spectral Characteristics Influenced by Both Contributing Factors}
\subsubsection{Flares with Successful Filament Eruption}
Figure \ref{fig_3} presents the imaging and spectral observations of Event 3, an M8.7 flare occurred on 2024 July 29 (see the online animation associated with Figure \ref{fig_3}). According to GOES observations, the flare began at 12:47 UT, peaked at 12:55 UT, and ended at 13:04 UT. The $\Delta EW$ shows similar feature as Event 1, peaking during the rise phase of the flare, before the peak of the GOES SXR flux and near its temporal derivative (see Figure \ref{fig_3}(d)). This event is accompanied by a successful filament eruption and an obvious CME (see Figure \ref{fig_a3}(b)). AIA 131 and 171 \AA\ images clearly show filament eruption structures (see Figures \ref{fig_a3}(a2), (a3), (c1), and (c2)). The filament erupted with significant velocity in both the line-of-sight and perpendicular directions, and did not obscure the flare ribbon region, so that we can clearly distinguish the spectral contribution from flare ribbons and the erupting filament by spatial decomposition. In this analysis, we examine three regions separately: the flare ribbon region, the filament propagation area, and the combined region that includes both.

In the flare ribbon region (indicated by cyan boxes in Figures \ref{fig_3}(a1)--(a3) and (b1)--(b3)), the $\Delta S(t,\lambda)$ shows line center emission, line broadening over $\pm$1.8 \AA, and red-asymmetry with redshifted velocity of $\sim+$44 km s$^{-1}$ (see Figure \ref{fig_3}(e2)), which is quiet similar to that for Event 1. However, due to the obvious filament upraising between 12:45 and 12:50 UT, the \saas\ line profile shows faint accelerated blueshifted absorption (see blue ellipse in Figure \ref{fig_3}(e2)). In the filament region (inside the green boxes but outside the cyan boxes in Figures \ref{fig_3}(a) and (b)), $\Delta S(t,\lambda)$ exhibits an obvious blueshifted absorption, together with emission near line center (see Figure \ref{fig_3}(e3)). The \saas\ spectral characteristics of filament region is quite similar to that of the successfully erupting region in Event 2. In this event, after rapidly accelerating to about -130 km s$^{-1}$, the blueshifted absorption gradually weakens and fades away in about 10 minutes (see the blue ellipse in Figure \ref{fig_3}(e3)). For the combined region, the $\Delta S(t,\lambda)$ shows emission near line center and in red wing, while shows prominent blueshifted absorption without a clear deceleration phase and the absence of redshifted absorption (see Figures \ref{fig_3}(c) and (e1)). One can see that the blueshifted absorption signals and the emission near the line center remain relatively complete within the wavelength range. As a result, the overall impact of the CHASE wavelength coverage limitation on this event is also minimal. The eruptive components can also be seen in EVE observations (see Figures \ref{fig_a3}(e) and (f)). We applied a double Gaussian fitting to the \ov\ line profile at its peak time, revealing a blueshifted emission component with a velocity of approximately $-$320 km s$^{-1}$. The \ov\ light curve reaches its maximum value near the peak time of \ha\ $\Delta EW$. The timings of blueshifted components in \ha\ and \ov\ match well (see Figure \ref{fig_11}(a)), while the \ov\ lines exhibit significantly higher velocities than those derived from \ha. These results also suggest that if similar spectra are observed in stellar flares, a stellar CME is very likely present.

In Event 3, the contributions of cold plasma in the erupting filament and hot plasma in the flare ribbons are both clearly visible in the \saas\ \ha\ spectra, as their signal strengths are of similar orders of magnitude. However, in most stellar observations, the flare energy is significantly higher, often resulting in much stronger spectral emission. To investigate how the temporal spectral characteristics change when the emission caused by flare ribbon dominates, we subsequently analyze an extremely strong solar flare event.

\begin{figure*}[htbp]
    \centering
    \includegraphics[width=0.9\textwidth]{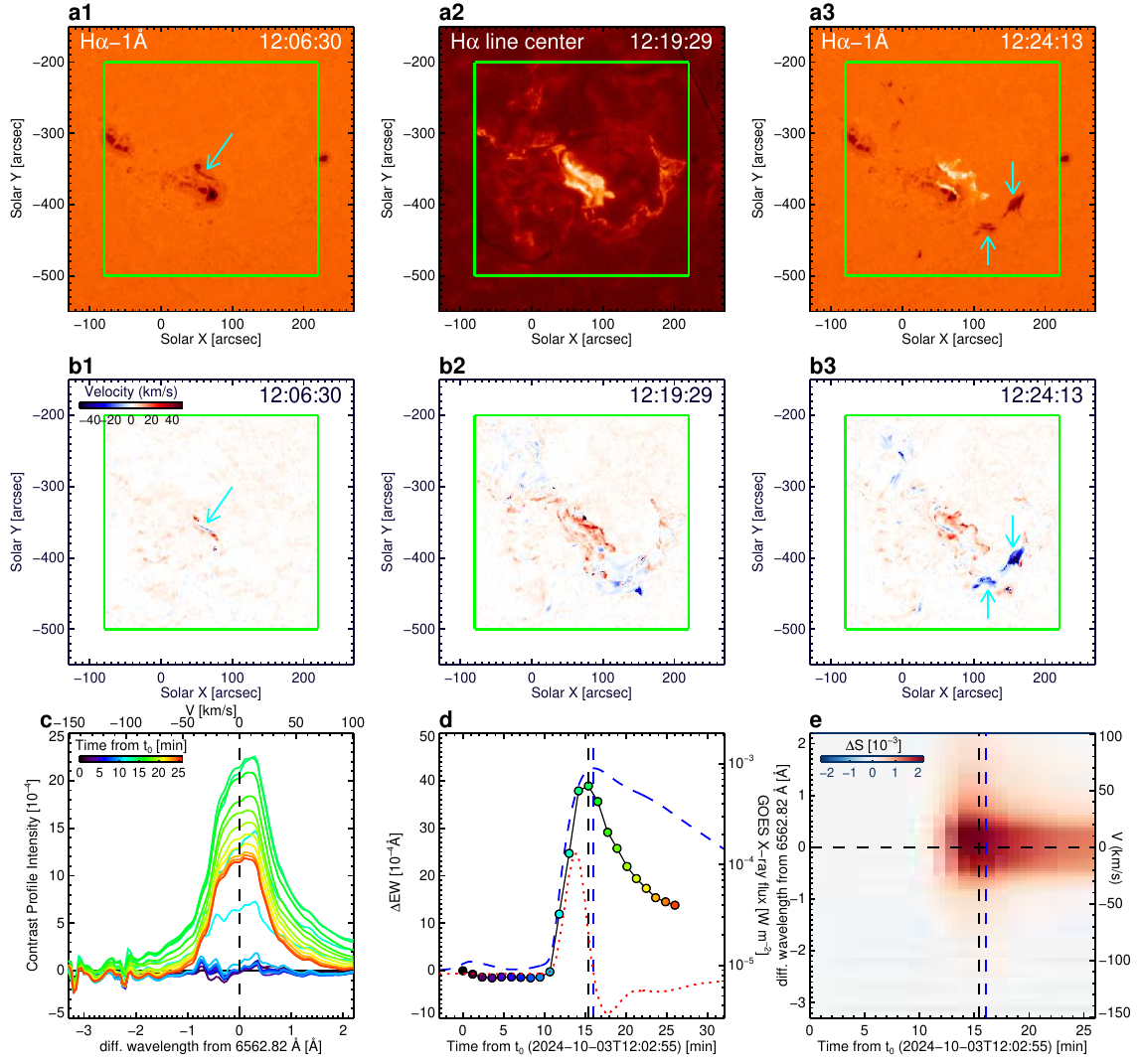}
    \caption{\chase~imaging and spectral observations of the X9.0 flare, occurred on 2024 October 03. (a1)--(a3): Sequence of CHASE \ha\ blue wing and line center images showing the source region (green boxes). The cyan arrow marks the regions with blueshifted velocity. (b1)--(e): The same as Figures \ref{fig_1} (b1)--(e) , but for this event. An animation (Figure7.mp4) covering 12:02 UT to 12:28 UT is available online, which displays the evolution of Event 4 through \ha\ $-$ 1 \AA\ , \ha\ center, and \ha\ $+$ 1 \AA\ intensity images. The animation’s duration is 3 seconds.}
    \label{fig_4}
\end{figure*}

\begin{figure*}[htbp]
	\centering
	\includegraphics[width=0.9\textwidth]{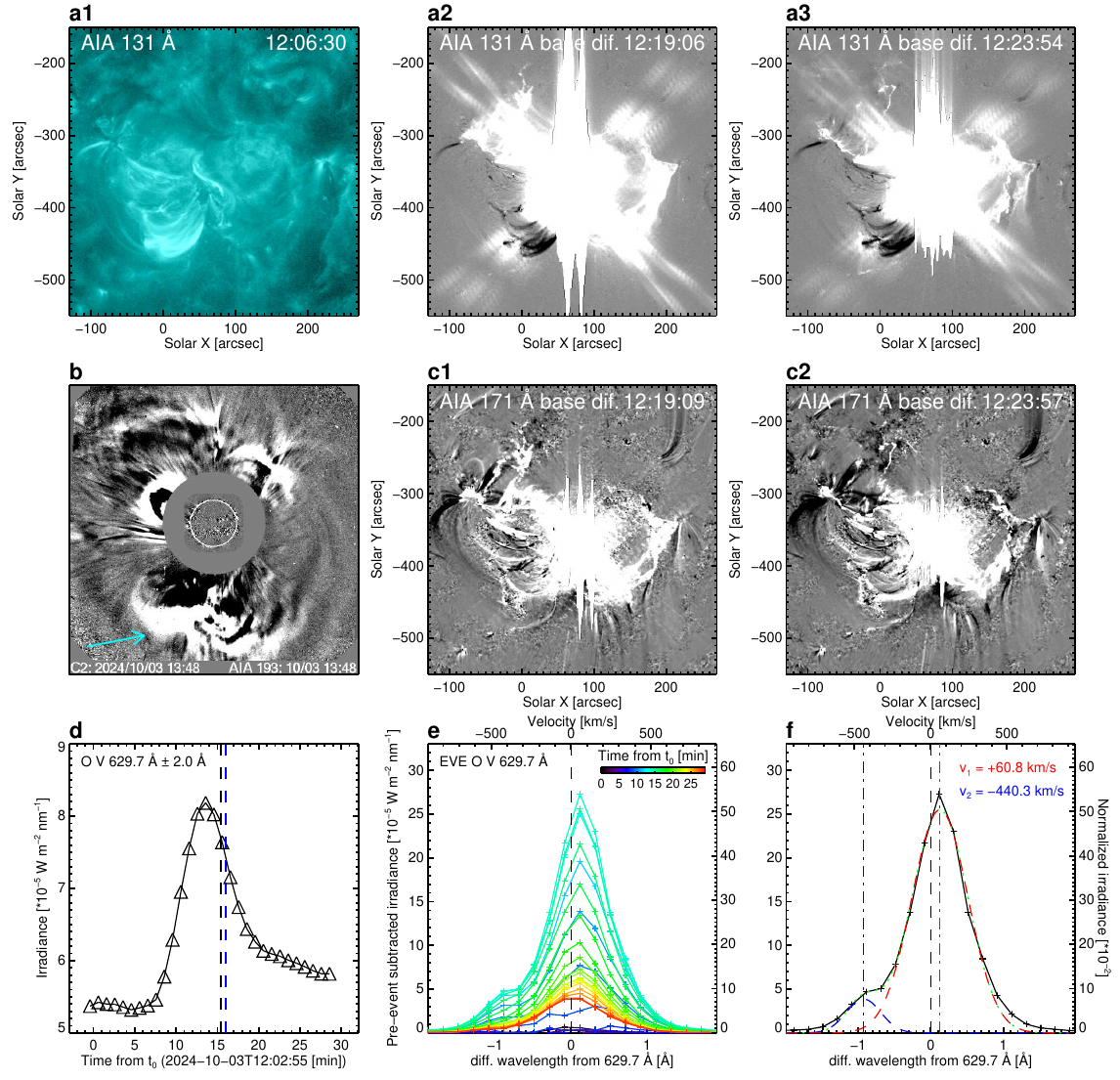}
	\caption{Multi-wavelength observations of the X9.0 flare, occurred on 2024 October 03. (a1)--(a3): AIA 131~\AA\ raw and running difference images taken at times corresponding approximately to those shown in panels (a1)--(a3) of Figure~\ref{fig_4}. (b): LASCO C2 running difference image at 13:48 UT, with a cyan arrow indicating the CME associated with this filament eruption event. (c1)--(c2): AIA 171~\AA\ running difference images taken at times corresponding approximately to those shown in panels (a2)--(a3) of Figure~\ref{fig_4}. (d): Irradiance variations of \ov\ 629.7 \AA. The vertical dashed lines mark the \ha\ (black) and GOES (blue) peak times. (e): Evolution of pre-event-subtracted O~V spectra during this event. (f): Double Gaussian fitting result of the \ov\ line profile near its peak time, showing distinct redshifted and blueshifted emission components. }
	\label{fig_a4}
\end{figure*}

Figure \ref{fig_4} presents the imaging and spectral observations of Event 4, an X9.0 flare occurred on 2024 October 3 (see the online animation associated with Figure \ref{fig_4}). This event is the strongest flare since solar cycle 25, and is accompanied by an obvious CME (see Figure \ref{fig_a4}(b)). Based on \ha\ imaging observations (Figures \ref{fig_4}(a1)--(a3)), the erupted filament is located just above the flare ribbons. Moreover, the filament eruption structure is not detectable in the AIA 131 and 171 \AA\ images (Figures \ref{fig_a4}(a2), (a3), (c1), and (c2)). As a result, it is difficult to fully distinguish the spectral contributions of the flare and the filament. Therefore, our analysis was conducted using the entire source region. According to GOES observations, the flare began at 12:08 UT, peaked at 12:18 UT, and ended at 12:27 UT. The $\Delta EW$ shows a significant peak near the peak of the GOES SXR flux (see Figure \ref{fig_4}(d)). In this event, $\Delta S(t,\lambda)$ of source region exhibits typical characteristic of flare ribbon region, including line center emission, line broadening over $\pm$1.4 \AA, and red asymmetry with the redshifted velocity of $\sim+$50 km s$^{-1}$ (see Figures \ref{fig_4}(c) and (e)). Although this event is accompanied by a prominent CME and obvious dark structures with blue velocity are visible in blue-wing images and Doppler maps (indicated by arrows in Figures \ref{fig_4}(a1), (a3), (b1), and (b3)), the \saas\ line profiles do not show blueshifted absorption. Similar to Event 1, because the emission signal near the line center is within the wavelength range, the limited wavelength coverage does not affect the analysis of this event as well. The peak value of $\Delta S(t,\lambda)$ in this event is on the order of $10^{-3}$, whereas in previous events, it ranged from $10^{-4}$ to $10^{-5}$. During violent flares like this event, it may be difficult if not impossible to determine the presence of CMEs only through the \ha\ line. In observations of stellar flares on solar-type stars, the detected super-flares are significantly more intense than this event (e.g., \citealt{2013ApJS..209....5S},  \citealt{2015EP&S...67...59M}). If these super-flares behave similarly to solar flares, identifying their corresponding CMEs using \ha\ line alone would be challenging. To track stellar CMEs, multi-wavelength spectral observations are needed \citep[e.g.,][]{2022ApJ...931...76X, 2023ApJ...953...68L, 2024ApJ...964...75O}. The \ov\ spectra show two distinct emission peaks (see Figures \ref{fig_a4}(e) and (f)), with a blueshifted velocity of $\sim-$440 km s$^{-1}$ and a redshifted velocity of $\sim+$60 km s$^{-1}$, which corresponding to CME and chromospheric condensation (\citealt{2019ApJ...875...93C}), respectively. The \ov\ light curve reaches its maximum during the flare's impulsive phase, which is earlier than the peak times of both \ha\ $\Delta EW$ and GOES SXR flux in this event. Although the blueshifted signal is barely detectable in the \ha\ spectra, it is clearly visible in the \ov\ lines (see Figure \ref{fig_11}(b)), reaching its max velocity before the peak time.

\begin{figure*}[htbp]
    \centering
    \includegraphics[width=0.9\textwidth]{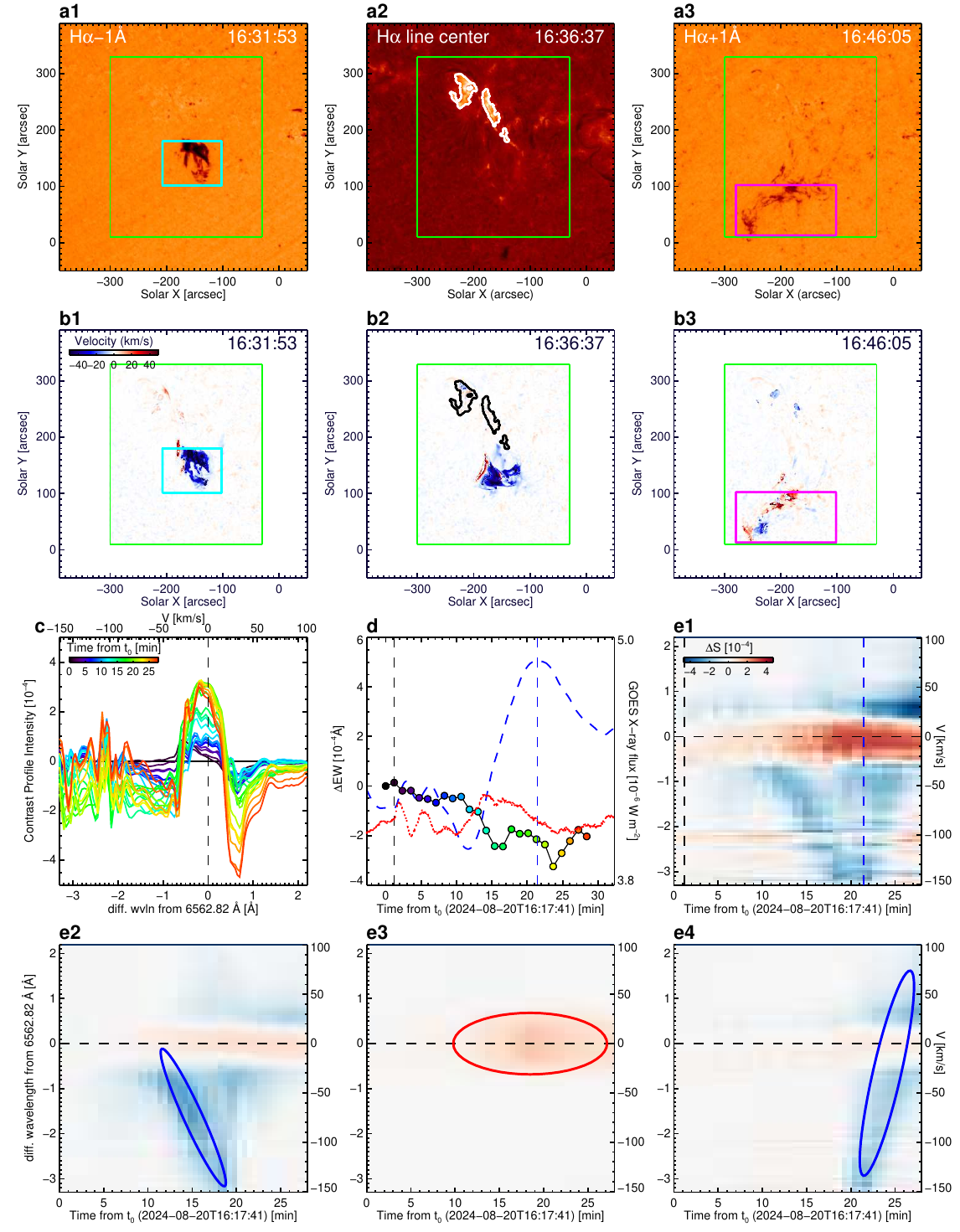}
    \caption{\chase~imaging and spectral observations of the C4.9 flare, occurred on 2024 August 20. (a1)--(a3): Sequence of CHASE \ha\ blue wing, line center, and red wing images showing the source region (green boxes). The cyan box, white/black contour and magenta box respectively indicate the regions with upwards moving plasma, flare ribbons, and downwards moving plasma. (b1)--(d): The same as Figures \ref{fig_1} (b1)--(d) , but for this event. (e1)--(e4): The time series of Sun-as-a-star \ha\ dynamic spectra of the green box, cyan box, white/black contour and magenta box, respectively. An animation (Figure9.mp4) covering 16:17 UT to 16:46 UT is available online, which displays the evolution of Event 5 through \ha\ $-$ 1 \AA\ , \ha\ center, and \ha\ $+$ 1 \AA\ intensity images. The animation’s duration is 3 seconds.}
    \label{fig_5}
\end{figure*}

\begin{figure*}[htbp]
	\centering
	\includegraphics[width=0.9\textwidth]{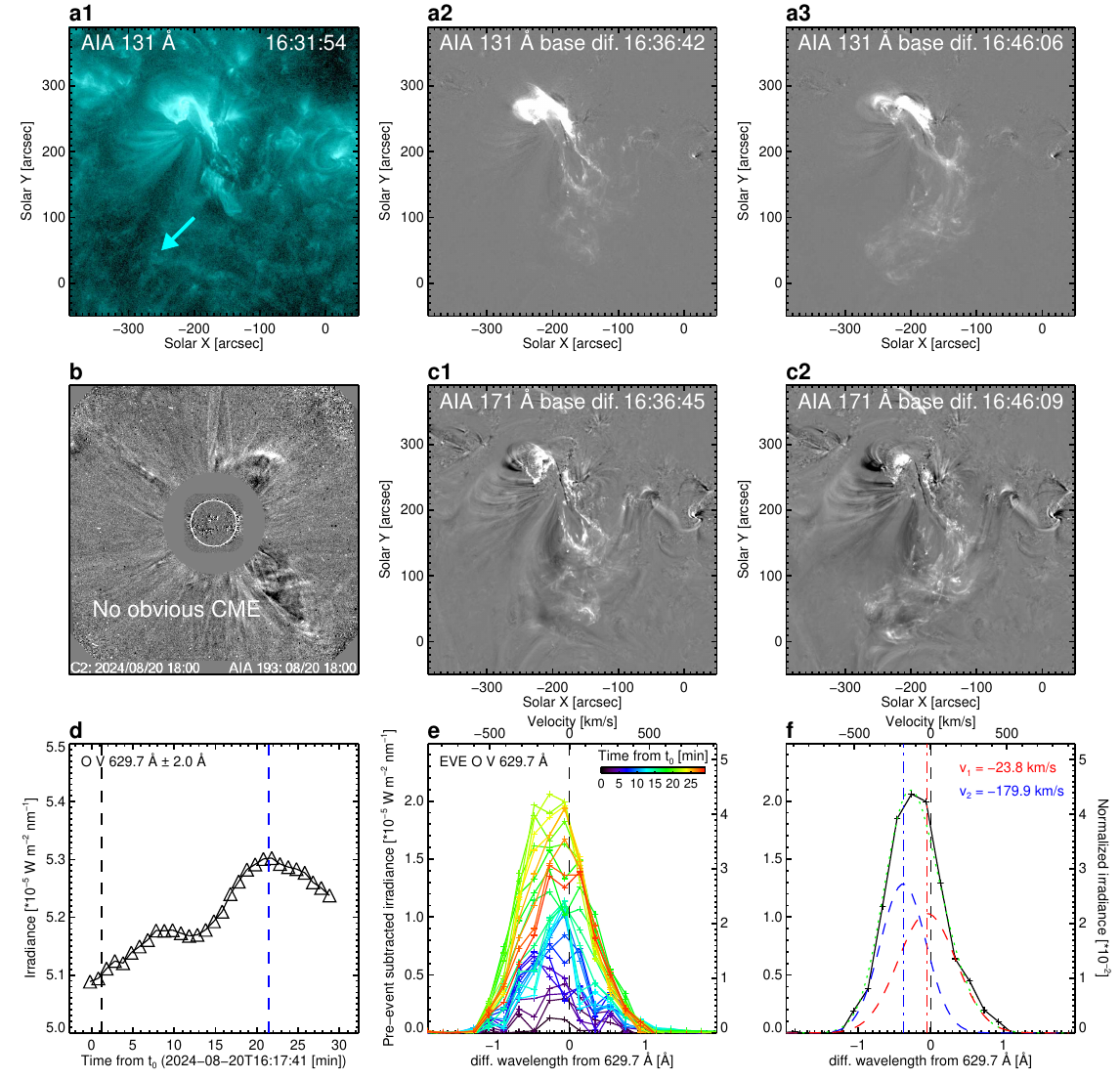}
	\caption{Multi-wavelength observations of the C4.9 flare, occurred on 2024 August 20. (a1)--(a3): AIA 131~\AA\ raw and running difference images taken at times corresponding approximately to those shown in panels (a1)--(a3) of Figure~\ref{fig_5}. (b): LASCO C2 running difference image at 18:00 UT. (c1)--(c2): AIA 171~\AA\ running difference images taken at times corresponding approximately to those shown in panels (a2)--(a3) of Figure~\ref{fig_5}. (d): Irradiance variations of \ov\ 629.7 \AA. The vertical dashed lines mark the \ha\ (black) and GOES (blue) peak times. (e): Evolution of pre-event-subtracted O~V spectra during this event. (f): Double Gaussian fitting result of the \ov\ line profile near its peak time, showing distinct central and blueshifted emission components.}
	\label{fig_a5}
\end{figure*}

\subsubsection{Flare with Failed Filament Eruption}
\begin{figure*}[htbp]
	\centering
	\includegraphics[width=0.98\textwidth]{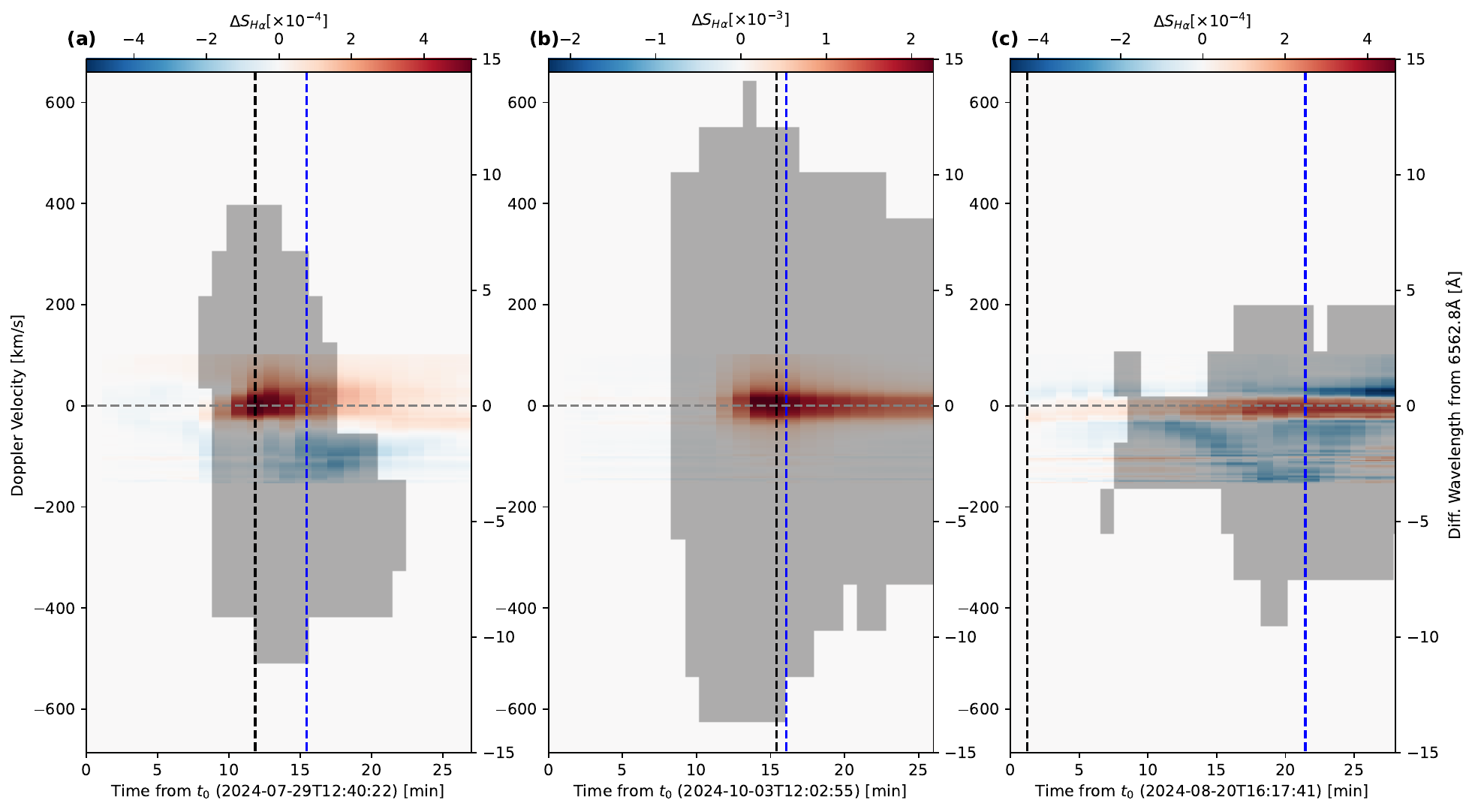}
	\caption{The composite dynamic spectra of \ha\ and \ov\ 629.7 \AA. (a): The composite dynamic spectra for Event 3. The \ha\ spectra are the same as that in Figure \ref{fig_3}, and the \ov\ spectra with absolute change in irradiance larger than 3$\sigma_{max}$ are shown in gray (see Section\ref{s33} for the details of $\sigma_{max}$). (b)--(c): The same as (a) but for Events 4 and 5, respectively.}
	\label{fig_11}
\end{figure*}

Figure \ref{fig_5} presents the imaging and spectral observations of Event 5, a C4.9 flare occurred on 2024 August 20 (see the online animation associated with Figure \ref{fig_5}). This flare is classified as a type-II confined flare (according to the result in \citealt{2019ApJ...881..151L}), which is accompanied by a failed filament eruption, and no CME was observed. AIA 131 and 171 \AA\ images reveal clear filament eruption structures (see Figures \ref{fig_a5}(a2), (a3), (c1) and (c2)), which fall back to the solar surface shortly after the eruption. Due to the significant lateral motion of the filament plasma, we can clearly distinguish its acceleration and deceleration phases through spatial decomposition. Accordingly, we analyze three sub-regions separately: the filament acceleration region, the flare ribbon region, and the filament deceleration and fallback region. According to GOES observations, the flare began at 16:28 UT, peaked at 16:39 UT, and ended at 16:59 UT. The $\Delta EW$ exhibits an overall dimming trend during this event, although a slight enhancement is observed near the peak time, but it remains lower than pre-event level (see Figure \ref{fig_5}(d)). In the filament acceleration region, the \saas\ \ha\ spectra show blueshifted absorption, with the velocity gradually increasing to approximately $-$150 km s$^{-1}$ near 16:36 UT (see blue ellipse in Figure \ref{fig_5}(e2)), corresponding to the upward acceleration of filament plasma. If the filament does not erupt successfully, the accelerated plasma will gradually decelerate and then accelerate downwards, and the $\Delta S(t,\lambda)$ shows blueshifted absorption with decreasing velocity and gradually turns to redshifted absorption (see blue ellipse in Figure \ref{fig_5}(e4)). The $\Delta S(t,\lambda)$ of the flare ribbon region shows line broadening over $\pm$0.9 \AA, emission at the line center, and redshifted velocity of about $+$20 km s$^{-1}$(see red ellipse in Figure \ref{fig_5}(e3)). For the entire source region, the \saas\ \ha\ spectra (see Figures \ref{fig_5}(c) and (e1)) exhibit certain similarities to those observed in Event 2, particularly in terms of overall spectral structure. However, the intensity of both the line center emission and the redshifted absorption components is stronger. Due to the limited wavelength coverage, $-$150 km s$^{-1}$ obtained around 16:36 UT in Figure \ref{fig_5} (e2) should be the lower limit value of the true maximum blueshifted velocity. However, as the deceleration phase begins shortly after the blueshifted absorption reaches the lower boundary of the spectral range, the overall impact of the CHASE wavelength coverage limitation on our analysis here remains minimal. The \ov\ spectra exhibit blueshifted emission with a velocity of around $-$180 km s$^{-1}$ (see Figures \ref{fig_a5}(e) and (f)). The \ov\ light curve does not exhibit dimming, and its peak time is close to that of the \goes\ SXR flux. The timings of blueshifted components of \ha\ and \ov\ also match well (see Figure \ref{fig_11}(c)), and their velocities are relatively comparable. In this event, prominent blueshifted features are observed both in \ha\ and EUV wavebands, but no CME is observed. This observation provides an important insight: the presence of a blueshifted component in a stellar spectrum—often used as a CME indicator—does not unequivocally imply the occurrence of a stellar CME. Therefore, CME diagnostics using spectral observations should be interpreted with caution.

\section{Conclusions}\label{s5}
In this work, we perform \saas\ analysis on five solar eruptive events with distinct characteristics. Although the five events differ from one another, our analysis reveals that the spatially-integral spectra of sub-regions, dominated by same physical process, exhibit similar temporal spectral characteristics. We also find that the present of blueshifted absorption features does not necessarily indicate the occurrence of a CME. However, we do identify a distinct spectral signature which is highly likely associated with CME events. Besides, some spectral signatures may be obscured when the magnitudes of emission and absorption are markedly different. Our main results are listed as follows:

1. By conducting a regional analysis, we can more precisely characterize the spectral signatures of different physical processes. The \saas\ \ha\ line profiles in flare ribbon regions exhibit line center emission, line broadening, and red asymmetry, which are associated with chromospheric condensation, consistent with previous works (\citealt{2022ApJ...933..209N}; \citealt{2022ApJ...939...98O}; \citealt{2024ApJ...966...45M}. In the evolution regions of erupting filaments, the \ha\ spectra display accelerated blueshifted absorption, decelerated blueshifted absorption, and redshifted absorption, capturing the complete evolution of cold plasma within filament moving upward and downward, consistent with previous studies (\citealt{2022NatAs...6..241N}; \citealt{2022ApJ...939...98O}). During the filament uplift phase, the \saas\ \ha\ spectra exhibit accelerated blueshifted absorption, corresponding to the upward acceleration of filament plasma. If the filament erupts successfully, this blueshifted absorption component gradually weakens and disappears as the ejected filament becomes optically thinner and eventually invisible in the \ha\ waveband. In contrast, for failed or partial eruptions, the \saas\ \ha\ spectra show blueshifted absorption with decreasing velocity and gradually turn to redshifted absorption, reflecting the deceleration and subsequent downward motion of the filament plasma. These results indicate that \saas\ spectra analysis in sub-regions of the eruption source region can better capture the characteristic temporal spectral features associated with different dynamical processes.

2. Through \ha\ spectral observations, an event with blueshifted absorption (Event 5), which is usually considered as a typical feature of CME, may not be accompanied by a CME. Conversely, an event without blueshifted absorption (Event 4) and an event with redshifted absorption (Event 2), which are often interpreted as lacking CMEs, may be associated with a CME. These mean that we should be more careful when interpreting the physical processes of stellar \ha\ spectra. However, we identify a distinct spectral signature highly likely associated with a flare accompanied by a CME (Event 3), characterized by the combination of emission near line center and in red wing, prominent blueshifted absorption without a clear deceleration phase, and the absence of redshifted absorption.

3. The emission from heated plasma in flare ribbons and the absorption from cold plasma in evolving filaments can be simultaneously observed in \saas\ \ha\ spectra only when their intensities are of comparable magnitude. Otherwise, some emission and absorption signals may be suppressed when their intensities differ greatly, which is also mentioned in previous work by \citet{2022ApJ...939...98O}. As shown in Event 4, only flare-related signatures are identified in the \saas\ \ha\ spectra, despite the occurrence of a violent CME. This is because the absorption produced by the cold filament plasma is significantly weaker than the intense emission from the heated flare ribbons and is therefore obscured in the spatially-integral spectra. A similar situation is likely in stellar observations. Most of the detected stellar events on solar-type stars are super-flares, which generate extremely strong spectral emission. Under such conditions, filament-related absorption features would be easily masked in stellar \ha\ spectra, unless there is a super-filament eruption associated with a super-CME. This may partially explain the relatively small number of confirmed stellar CMEs.

4. Through checking EVE observation of \ov\ spectral line, formed at a higher temperature than \ha, we find that in Events 3, 4, and 5, the Sun-as-a-star \ov\ spectra all exhibit blueshifted emission features. Notably, the blueshifted velocities derived from the OV lines in Events 3 and 4, both associated with CMEs, are significantly higher than those inferred from the corresponding \ha\ spectra. In contrast, in Event 5, which is not accompanied by a CME, the blueshifted velocities in both \ov\ and \ha\ lines are relatively comparable. These results indicate that the combination and comparisons of \ha\ and UV spectral observations can effectively reveal the velocity evolution of erupting filaments and potential existence of associated CMEs, depicting a more comprehensive scenario of the solar and stellar filament eruption.

In this study, the measured line broadening widths and redshifted velocities of \ha\ spectra in solar eruption source regions, contributed by hot plasma in flare ribbons, around peak time are approximately in range of $\pm$0.9 to $\pm$1.8 \AA\ and $+$20 to $+$50 km s$^{-1}$, respectively. These values are quite lower than those reported in stellar flares studies \citep[e.g.,][]{2022ApJ...928..180W, 2024ApJ...961..189N}. Additionally, the maximum blueshifted velocities of cold plasma within filaments, as indicated by \ha\ line profiles, reach about $-$130--$-$150 km s$^{-1}$, consistent with that of previous solar flare studies \citep[e.g.,][]{2022ApJ...939...98O, 2022ApJ...933..209N}.

As demonstrated in this work, spatially-integral \ha\ and EUV spectra are highly useful for studying stellar active events but still have certain limitations. In the future, we will analyze a broader range of solar active events that involve each physical processes revealed in this work and find the temporal spectral characteristics of different combinations of these processes, using multi-wavelength observations for more comprehensive diagnostics. Additionally, we will examine time-resolved spectral observations of stellar eruptive events to identify features similar to those observed in solar eruptions, which will help diagnose the plasma dynamic processes occurring in the source regions of stellar eruptive events.

\section*{ACKNOWLEDGMENTS}
The authors appreciate the anonymous referee for the constructive comments and valuable suggestions. The data are used courtesy of CHASE, SDO, SOHO, and GOES science teams. SDO is a mission of NASA’s Living With a Star Program. The CHASE mission is supported by the China National Space Administration (CNSA). The authors are supported by the Strategic Priority Research Program of the Chinese Academy of Sciences (XDB0560000), the National Key R\&D Program of China (2022YFF0503800 and 2022YFF0503004), the National Natural Science Foundation of China (12273060, 12222306, 12333009, and 12273115), the Youth Innovation Promotion Association CAS (2023063), postgraduate Research \& Practice Innovation Program of Jiangsu Province (KYCX24\_0183), and the Specialized Research Fund for State Key Laboratory of Solar Activity and Space Weather.

\bibliographystyle{apj}

\end{document}